\documentstyle[floats,eqsecnum,multicol,aps,psfig]{revtex}

\def\Sv{\vec{S}}
\def\Jt#1{\tilde{J}_{#1}}
\def\Jb#1{\bar{J}_{#1}}
\def\Jc#1{{\cal J}_{#1}}
\def\tr{{\rm tr}}
\def\Mag{\langle M \rangle}
\def\abs#1{\vert #1 \vert}
\def\beq{\begin{equation}}
\newcommand{\eeq}[1]{\label{#1} \end{equation}}
\def\bea{\begin{eqnarray}}
\def\eea{\end{eqnarray}}
\def\nn{\nonumber}
\def\vec#1{{\bf #1}}

\def\breakon{\end{multicols}\vspace{-.7cm}
\noindent\rule{.49\linewidth}{.3mm}\rule{.3mm}{.5cm}\vspace{0.0cm}}
\def\breakoff{\vspace{-0.25cm}
\noindent
\rule{.50\linewidth}{.0mm}\rule[-.47cm]{.3mm}{.5cm}\rule{.49\linewidth}{.3mm}
\vspace{-0.55cm}
\begin{multicols}{2}   }
\begin{document}

\makeatletter
\renewenvironment{table}
  {\let\@capwidth\linewidth\def\@captype{table}}
  {}

\renewenvironment{figure}
  {\let\@capwidth\linewidth\def\@captype{figure}}
  {}
\makeatother

\draft

\title{Frustrated trimer chain model and
Cu$_3$Cl$_6$(H$_2$O)$_2\cdot$2H$_8$C$_4$SO$_2$ in a magnetic field}
\author{A.\ Honecker$^1$ and A.\ L\"auchli$^2$}
\address{$^1$Institut f\"ur Theoretische Physik, TU Braunschweig,
           Mendelssohnstr.\ 3, D--38106 Braunschweig, Germany. \\
        $^{2}$Institut f\"ur Theoretische Physik, ETH-H\"onggerberg,
           CH--8093 Z\"urich, Switzerland.}

\date{May 24, 2000; revised October 13, 2000}
\maketitle
\begin{abstract}
\begin{center}
\parbox{14cm}{Recent magnetization and
susceptibility measurements on Cu$_3$Cl$_6$(H$_2$O)$_2\cdot$2H$_8$C$_4$SO$_2$
by Ishii {\it et.al.}\ [J.\ Phys.\ Soc.\ Jpn.\ {\bf 69}, 340 (2000)]
have demonstrated the existence of a spin gap. In order to explain the
opening of a spin gap in this copper-trimer system, Ishii {\it et.al.}\
have proposed a frustrated trimer chain model. Since the exchange constants
for this model have not yet been determined, we develop a twelfth-order
high-temperature series for the magnetic susceptibility and fit it
to the experimentally measured one. We find that some of the coupling
constants are likely to be {\it ferromagnetic}. The combination of several
arguments does not provide any evidence for a spin gap in the parameter
region with ferromagnetic coupling constants, but further results
e.g.\ for the magnetization process are in qualitative agreement with the
experimental observations.}
\end{center}
\end{abstract}


\begin{multicols}{2}

\section{Introduction}

The trimerized $S=1/2$ Heisenberg chain in a strong external magnetic
field has already received a substantial amount of theoretical
attention, one reason being a plateau at one third of the
saturation magnetization in the magnetization curve
\cite{Hida,Okamoto,CaGy,poly,OkKi}. Some frustrated variants of
the trimer model have also been investigated \cite{SuSha,LoFe,LoSi,Takano,TKS}
since they can be shown to have dimer groundstates and thus a
spin gap.

While many materials with trimer constituents exist
(see e.g.\ \cite{BWRZZHD}), the behavior in high magnetic fields
has been investigated only in a few of them, for instance in
3CuCl$_2 \cdot$2dioxane \cite{AAIAG}. Also
Cu$_3$Cl$_6$(H$_2$O)$_2\cdot$2H$_8$C$_4$SO$_2$ belongs to
the known trimer materials \cite{SwWi,SLW}, but its behavior
in a strong magnetic field has been measured only recently \cite{ITHUOTKN}
and at the same time its magnetic susceptibility has been remeasured.
Surprisingly, a spin gap of about 3.9Tesla (that is roughly 5.5K) is observed
both in the magnetic susceptibility of
Cu$_3$Cl$_6$(H$_2$O)$_2\cdot$2H$_8$C$_4$SO$_2$ \cite{footnote1}
as well as in the magnetization as a function of external magnetic field.
This system probably exhibits also a plateau at one third of the saturation
magnetization in addition to the spin gap.

\begin{figure}
\centerline{
\setlength{\unitlength}{0.00041700in}%
\begin{picture}(7330,2530)(1736,-5226)
\thicklines
\put(1781,-5141){\line( 1, 1){2400}}
\put(1791,-5151){\line( 1, 1){2400}}
\put(1801,-5161){\line( 1, 1){2400}}
\put(4181,-5141){\line( 1, 1){2400}}
\put(4191,-5151){\line( 1, 1){2400}}
\put(4201,-5161){\line( 1, 1){2400}}
\put(6581,-5141){\line( 1, 1){2400}}
\put(6591,-5151){\line( 1, 1){2400}}
\put(6601,-5161){\line( 1, 1){2400}}
\thinlines
\put(1801,-2761){\line( 1, -1){2400}}
\put(4201,-2761){\line( 1, -1){2400}}
\put(6601,-2761){\line( 1, -1){2400}}
\thicklines
\multiput(4201,-2761)(0,-100){24}{\line(0,-1){50}}
\multiput(6601,-2761)(0,-100){24}{\line(0,-1){50}}
\put(4251,-4141){\makebox(0,0)[lb]{\smash{\Large $J_2$}}}
\put(4801,-3361){\makebox(0,0)[lb]{\smash{\Large $J_1$}}}
\put(4801,-4861){\makebox(0,0)[lb]{\smash{\Large $J_3$}}}
\end{picture}}
\medskip

\caption{The frustrated trimer chain model. All corners and intersections
carry a spin 1/2 coupled with exchange constants indicated by the connecting
lines.
\label{Model}
}
\end{figure}
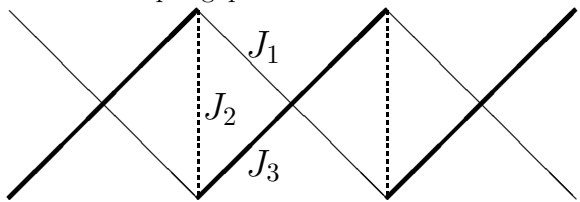

Motivated by the crystal structure \cite{SwWi}, the authors of
\cite{ITHUOTKN} have proposed the following model
(see also Fig.\ \ref{Model}) \cite{footnote2}:
\bea
H &=& J_{1} \sum_{i=1}^{L/3} \left\{ \Sv_{3 i} \cdot \Sv_{3 i + 1}
          + \Sv_{3 i+1} \cdot \Sv_{3 i + 2} \right\} \nn \\
&&+ J_{2} \sum_{i=1}^{L/3} \Sv_{3 i + 2} \cdot \Sv_{3 i + 3} \nn \\
&&+ J_{3} \sum_{i=1}^{L/3} \left\{ \Sv_{3 i + 1} \cdot \Sv_{3 i + 3}
          + \Sv_{3 i + 2} \cdot \Sv_{3 i + 4} \right\} \nn \\
&& - h \sum_{i=1}^{L} S^z_i \, .
\label{Htrimer}
\eea
Since the spin is localized on Cu$^{2+}$ ions, the
$\Sv_i$ are spin-1/2 operators at site $i$. In (\ref{Htrimer}),
the reduced field $h$ is related to the physical field $H$
by $h = g \mu_B H$ in units where $k_B = 1$. The numerical prefactor
is determined by $\mu_B \approx 0.67171$K/Tesla as well as the value of
$g$ which for the present material is slightly above 2 (the precise
numerical value depends on the direction of the external magnetic field 
relative to the crystal axes).

For a study of the phase diagram of the Hamiltonian (\ref{Htrimer}),
it is useful to observe that the Hamiltonian and therefore also the
phase diagram are invariant under the exchange of $J_1$ and $J_3$ such
that one can concentrate e.g.\ on $\abs{J_3} \le \abs{J_1}$. In fact,
the $h=0$ phase diagram with antiferromagnetic exchange constants ($J_i \ge 0$)
has been explored in \cite{OTTK} using bosonization and exact
diagonalization (see also \cite{SaTa}) determining in particular
a parameter region with a spin gap. Very recently, this was complemented
by a computation of the magnetization curve at some values of the
parameters using DMRG \cite{TOHTK}.
The investigations of \cite{OTTK,SaTa,TOHTK} concentrated on the region
with all coupling constants in (\ref{Htrimer}) antiferromagnetic
($J_i \ge 0$) because \cite{ITHUOTKN} suggested that this should
be appropriate for Cu$_3$Cl$_6$(H$_2$O)$_2\cdot$2H$_8$C$_4$SO$_2$.
However, the parameters relevant to the experimental system have
not really been determined so far. We believe that this is an important
issue in particular in view of the fact that
according to the crystallographic data \cite{SwWi},
all angles of the Cu--Cl--Cu bonds lie in the region of
$91^\circ$ to $96^\circ$ -- a region where usually no safe
inference on the coupling constants $J_i$ can be made, not even
about their signs. We will therefore develop a high-temperature
series for the magnetic susceptibility of the model (\ref{Htrimer}) and
use it to {\it determine} the coupling constants from the experimental
data \cite{ITHUOTKN}. It will turn out that some coupling constants
are likely to be {\it ferromagnetic}, {\it i.e.}\ the experimentally
relevant coupling constants lie presumably outside the region studied so far.
We then proceed to study more general properties of the model and to
address the question of a spin gap in the
relevant parameter region. We use mainly perturbative arguments
supplemented by numerical methods.

Some supplementary results on the trimer model
are contained in the appendices or can be found in \cite{WWW}.

\section{Magnetic susceptibility and specific heat}

\subsection{High-temperature series for zero field}

First we discuss some high-temperature series in zero magnetic field.
We have used an elementary approach to perform the computations.
Denote the Hamiltonian of a length $L$ chain with $h=0$ by $H_0$.
Then the fundamental ingredient for any higher-order expansion
is that contributions of $\tr\left(H_0^N\right)$ to suitable physical quantities
become independent of the system size $L$ if one uses a long enough
chain with periodic boundary conditions. The concrete Hamiltonian
$H_0$ given by (\ref{Htrimer}) must be applied $2 L / 3$
times to wind once around the system and to feel that it is finite.
On the other hand, contributions from $\tr\left(H_0^N\right)$ with
$N < 2 L / 3$ are independent of $L$. We have used this observation
to determine the high-temperature series by simply computing the traces
for the lowest powers $N$ on a chain with
a fixed $L$ and periodic boundary conditions
\cite{footnote3}.
Just two small refinements
to this elementary approach have been made. The first one
is that we computed the traces separately for all subspaces
of the $z$-component of the total spin $S_{\rm tot}^z$. This
is already sufficient to obtain series for the specific heat
$c_v$ and the magnetic susceptibility $\chi$. The second one is
to make also the order $2 L / 3$ usable: At this order, only the
coefficient of $J_1^{L/3} J_3^{L/3}$ is affected by the finiteness
of the chain and this coefficient can be corrected by hand using
results for a Heisenberg ring of length $2 L /3$.

For notational convenience, we introduce the partition function for
$L$ sites by
\beq
Z_L = \tr\left({\rm e}^{-\beta H_0} \right)
\eeq{ZL}
with $k_B T = 1/\beta$.

\breakon

The lowest orders of a reduced magnetic susceptibility $\chi$ are found
to be
\bea
\chi_{\rm red.}(\beta) &=&
   {1 \over \beta L Z_L} {\partial^2 \over \partial h^2} \left.
\tr \left({\rm e}^{-\beta \left(H_0 - h S_{\rm tot}^z\right)} \right)
\right\vert_{h=0}
= {\beta \over L} {\tr\left(\left({S_{\rm tot}^z}\right)^2
  {\rm e}^{-\beta H_0} \right) \over Z_L} \nn \\
&=&
{\beta \over 4}
-{{\beta}^{2} \over 24} \left (2 J_3 + J_2 + 2 J_1\right )
-{\frac {{\beta}^{3}}{96}} \left (J_1^{2} + J_2^{2}+J_3^{2}
    -6 J_3 J_1-2 J_2 J_1 -2 J_3 J_2\right ) \nn \\
&&+ {\frac {{\beta}^{4}}{1152}} \left (8 J_3^{3}-3 J_3^{2} J_2
    +6 J_2^{2} J_1+6 J_3 J_2^{2}+8 J_1^{3}-3 J_2 J_1^{2}+ J_2^{3}
\right ) \nn \\
&&+ {\frac {{\beta}^{5}}{4608}} \left (-28 J_3^{2} J_2 J_1
    -28 J_3 J_1^{2} J_2+36 J_3^{2} J_1^{2}+8 J_3^{2} J_2^{2}
    +8 J_2^{2}J_1^{2}-2 J_2^{3} J_1-34 J_3 J_1^{3}-34 J_3^{3} J_1
\right.  \nn \\ && \qquad \left.
    -10 J_2 J_1^{3}-2 J_3 J_2^{3}-10 J_3^{3} J_2 -28 J_3 J_1 J_2^{2}
    +14 J_1^{4}+14 J_3^{4}+5 J_2^{4}\right ) \nn \\
&& +{\cal O}\left(\beta^6\right) \, .
\label{chiHT}
\eea
Similarly, we obtain the lowest orders of the high-temperature series
for the specific heat
\bea
{c_v(\beta) \over k_B} &=&
{\beta^2 \over L} {\partial^2 \over \partial \beta^2}
\ln\left( Z_L \right) \nn \\
&=& {{\beta}^2 \over 16} \left(2 J_1^{2}+ J_2^{2}+2 J_3^{2}\right )
+{{\beta}^{3} \over 32}
   \left(2 J_3^{3}+2 J_1^{3}+ J_2^{3} -6 J_3 J_1 J_2\right ) \nn \\
&& -{\frac {{\beta}^{4}}{256}} \left (8 J_3^{2} J_2 J_1 +8 J_3 J_1^{2} J_2
    +12 J_3^{2} J_1^{2}+8 J_3^{2} J_2^{2} +8 J_2^{2} J_1^{2}+6 J_1^{4}
    +6 J_3^{4}+ J_2^{4}+8 J_3 J_1 J_2^{2}\right ) \nn \\
&& +{\cal O}\left(\beta^5\right) \, .
\label{cvHT}
\eea
Complete 12th order versions of both series can be accessed via
\cite{WWW}.

For a uniform Heisenberg chain ($J_1 = J_2$ and $J_3 = 0$), the coefficients
of the series for $\chi$
and $\ln\left( Z_L \right)/L$ (or $c_v$) agree with those given for instance in
\cite{hTser} when they overlap.

\breakoff

\subsection{Fit to the experimental susceptibility}

\label{secChiFit}

Now we use our 12th order series for the susceptibility (\ref{chiHT})
to fit the experimental data \cite{ITHUOTKN} and thus extract values
for the coupling constants $J_i$. We used the data for the
single crystal ($H\parallel b$-axis) and the polycrystalline sample
\cite{ITHUOTKN} as well as some unpublished new measurements for all
three axes of a single crystal \cite{unpub}.
For the polycrystalline case the average $g$-factor is known to be
$g_{\rm av} \approx 2.1$ from ESR while in the single crystal case
\cite{ITHUOTKN} we used the $g$-factor as a fitting parameter.
The following prefactors are used to match the series (\ref{chiHT})
to the experimental data:
\beq
\chi_{\rm exp.}(T)=\frac{3 N_A g^2
\mu_B^2}{k_B}\ \chi_{\rm red.}\left({1 \over T}\right) \, .
\eeq{chiExp}
We performed fits in various intervals of temperature with a lower boundary
($T_l$) lying between 150K and 250K, while the upper boundary was kept fixed at
300K. Fits were performed with the raw 12th order series.
For both experimental data sets of \cite{ITHUOTKN} we obtained reasonable,
though volatile fits around $T_l = 150\mbox{K}-250\mbox{K}$ yielding the
following estimates: $J_1 = -250 \mbox{K} \pm 40\mbox{K}$,
$J_2 = 250 \mbox{K} \pm 40\mbox{K}$, $J_3= -40 \mbox{K} \pm 30 \mbox{K}$.
For the single crystal sample we additionally determined $g_b = 1.95 \pm 0.05$.

We have further performed fits to unpublished single-crystal data sets
where the $g$-factors are known from ESR \cite{unpub}. When a constant is
added to (\ref{chiHT}), the data for all three crystal axes can be fitted
consistently with $J_1 \approx -300$K, $J_2 \approx 280$K and
$J_3 \approx -60$K in an interval of high temperatures
($220{\rm K} \lesssim T \le 300$K).
This set of coupling constants is in agreement with our earlier fits and
we will use the latter in the further discussion below. 

\begin{figure}
\centerline{\psfig{figure=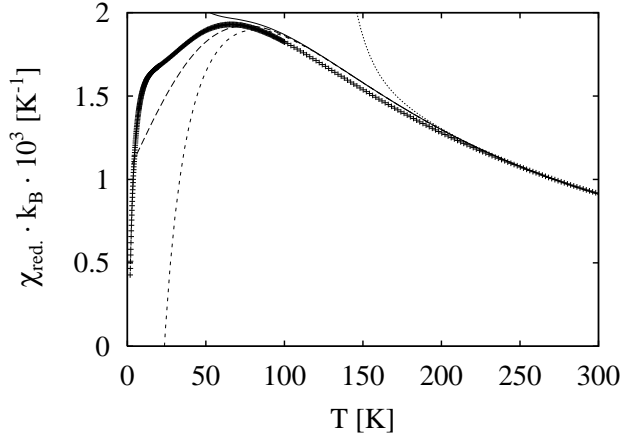,width=\linewidth,angle=270}}
\caption{
Experimental results for the susceptibility (`$+$') in comparison
with the fit $J_1 = -300$K, $J_2 = 280$K and $J_3 = -60$K.
We show the raw 12th order series (dotted line) as well as several
Pad\'e approximants: [7,6] (full line), [6,6] (long dashes) and
[6,5] (short dashes).
\label{chiFig}
}
\end{figure}

Fig.\ \ref{chiFig} shows the measured susceptibility for the
polycrystalline sample \cite{ITHUOTKN} together with the series
result. Since the parameters were obtained from a fit which was
performed with a different data set, we have used $g = 2.03$
(which differs slightly from the experimentally found $g_{\rm av} \approx 2.1$)
in order to obtain agreement of the raw series with the experimental
data for $T \ge 240$K. Clearly, the raw series should not be trusted
down into the region of the maximum of $\chi$ where Pad\'e approximants should
be used instead. The region below the maximum cannot be expected
to be described with a high-temperature series. The overall
agreement is reasonable though the theoretical result reproduces the
experimental one in the vicinity of the maximum only qualitatively.
This discrepancy might be due to the frustration in the
model which leads to cancellations in the coefficients.
Note also that, due to the frustration, the maximum of $\chi$ is
located at a lower temperature than would be expected for a
non-frustrated model with coupling constants of the same order
of magnitude. Consequently, higher orders are important in the
entire temperature range covered by Fig.\ \ref{chiFig}, precluding
in particular the analysis of the high-temperature tail of $\chi$
in terms of a simple Curie-Weiss law. 

The agreement for intermediate temperatures can be improved
if the maximum is included in the fitting region and
Pad\'e approximants are used in the fit. The main change with respect
to the fits discussed above is that $J_3$ tends to be closer to $J_1$.
However, it will become clear from the discussion in later sections that
the region with $J_3$ close to $J_1$ is not appropriate to describe the
experimental observations of the low-temperature region.

Although we are not able to determine the coupling constants to
high accuracy, all our fits lead to the conclusion that $J_2$ should be
antiferromagnetic and $J_1$ and $J_3$ (or at least one of them) must be
{\it ferromagnetic} if one wants to model the susceptibility measured at high
temperatures \cite{ITHUOTKN} with the frustrated trimer chain (\ref{Htrimer}).
In view of earlier theoretical investigations \cite{OTTK,SaTa,TOHTK},
this conclusion is somewhat surprising. Note that none of our fits
converged to all $J_i > 0$. Additional assumptions (including a constraint
on the $J_i$) are necessary to determine from $\chi(T)$ what the optimal
values of the $J_i$ would be in this antiferromagnetic region and thus
allow for a comparison with Fig.\ \ref{chiFig}. Such a fit and a comparison
with the present one is discussed in appendix \ref{appAF}. The upshot is
that the experimentally observed $\chi(T)$ \cite{ITHUOTKN} cannot be
explained with only antiferromagnetic $J_i$.

The findings of this section necessitate a detailed
re-analysis of the Hamiltonian (\ref{Htrimer}) since earlier works
did not look at the appropriate parameter region.

\section{Lanczos results}

In order to study the zero-temperature behavior of the frustrated
trimer chain we have performed Lanczos diagonalizations of small clusters
with periodic boundary conditions. Although computations were
performed for various values of the parameters, we will present explicit
results only for the final parameter set determined above. Further
results in the region $J_i > 0$ are in agreement with \cite{OTTK,SaTa,TOHTK}
and are used in appendix \ref{appAF}.

Fig.\ \ref{mCurve} presents the zero-temperature magnetization curve
for the trimer chain model.
Here and below the magnetization $\Mag$ is normalized to saturation
values $\pm 1$.
First, it is reassuring that the system still has antiferromagnetic
features despite two ferromagnetic coupling constants (note that
we are now probing a region far from that used for determining the
$J_i$). Since experiments found a spin gap \cite{ITHUOTKN}, an important
question clearly is if we also obtain a gap from the model with these
parameters. We have therefore performed a finite-size
analysis of the gap to $S^z = 1$ excitations (corresponding to the
first step in the finite-size magnetization curves of Fig.\ \ref{mCurve}).
All our approaches led to results compatible with a vanishing gap. However,
it is difficult to reliably exclude a gap of a few K with system sizes
$L \le 30$. We will therefore return to this issue later and assume for
the extrapolated thick line in Fig.\ \ref{mCurve} a vanishing spin gap.
In general, this extrapolation was obtained by connecting the mid-points
of the steps of the $L=30$ magnetization curve, except for $\Mag = 1$
and $\Mag = 1/3$ where the corners were used.

\begin{figure}
\centerline{\psfig{figure=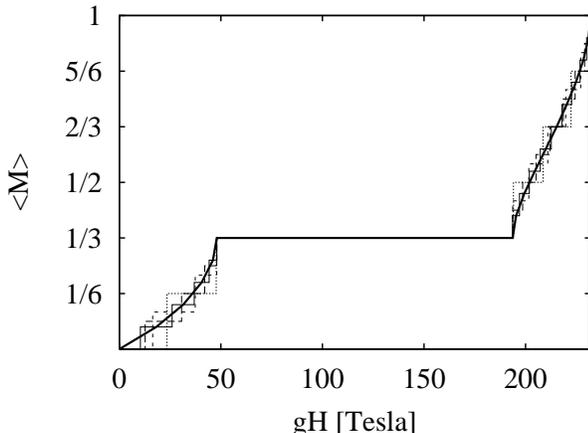,width=\linewidth,angle=270}}
\caption{
Magnetization curve for $J_1 = -300$K, $J_2 = 280$K and $J_3 = -60$K.
The thick line is an extrapolation whereas thin lines are for finite
system sizes: $L=12$ (dotted), $L=18$ (short dashes), $L=24$ (long dashes)
and $L=30$ (full).
\label{mCurve}
}
\end{figure}

For the parameters of Fig.\ \ref{mCurve}, $\Mag = 1/3$ is reached with
a magnetic field $H = 20-25$Tesla. The order of magnitude agrees
with the experimental finding \cite{ITHUOTKN} even if the value
found within the model is a factor of two to three below the
experimental one. Above this field,
Fig.\ \ref{mCurve} exhibits a clear $\Mag = 1/3$ plateau
which is expected on general grounds \cite{Hida,Okamoto,CaGy,poly,OkKi}.

\begin{figure}
\centerline{\psfig{figure=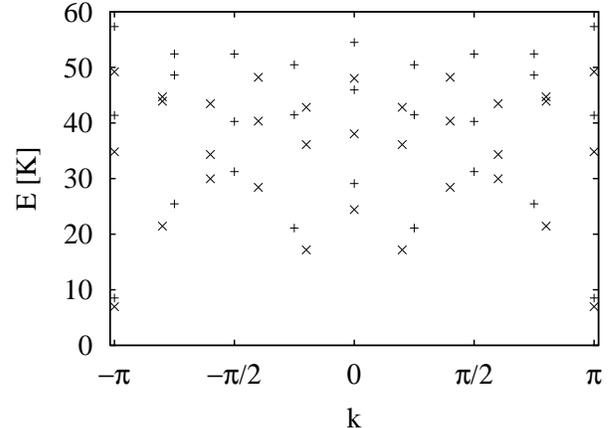,width=\linewidth,angle=270}}
\caption{
Lowest three excitations in the $S^z = 1$ sector for $L=24$
(`$+$') and $L=30$ (`$\times$') with $J_1 = -300$K, $J_2 = 280$K and
$J_3 = -60$K.
\label{dispFig}
}
\end{figure}

We conclude this section by presenting in
Fig.\ \ref{dispFig} the lowest three excitations for
the $S^z = 1$ sector as a function of momentum $k$, where
$k$ is measured with respect to the groundstate, {\it i.e.}\ $k = k_{S^z = 1}
- k_{\rm GS}$. This spectrum is very similar to that of an
$S=1/2$ Heisenberg chain of length $L/3$ with coupling
constant $J_{\rm eff.} \approx 16$K. In particular, one
can recognize the two-spinon scattering continuum
and a few higher excitations.
This identification of the low-energy excitations of the frustrated trimer
chain with an effective $S=1/2$ Heisenberg chain is one of the numerical
indications for the absence of a spin gap.

\section{The line $J_1 = J_3$}

\label{secTheLine}

The Lanczos results of the previous section raise the question
if the trimer chain model has a spin gap in the region with $J_1$,
$J_3 < 0$: In this region, the model behaves like an antiferromagnet
(see e.g.\ Fig.\ \ref{mCurve}) which is frustrated since the number
of antiferromagnetic coupling constants around a triangle is odd.
Therefore, a spin gap appears possible in principle and 
we proceed with further arguments to decide whether it appears in
the relevant parameter region.

Evidence for a spin gap in the parameter
region $J_1$, $J_3 > 0$ was actually first obtained on the line
$J_1 = J_3$ \cite{TKS}. The reason is presumably that
the line $J_1 = J_3$ can be treated analytically at least to some extent
because then the total spin is locally conserved on each bond coupled by $J_2$.
In fact, one can easily discuss the entire magnetization process \cite{HMT}
and not just the question of a spin gap and we refer the interested
reader to \cite{WWW} for some comments on this aspect.

Recall that the model (\ref{Htrimer}) with $h=0$
gives rise to three types of groundstates in different regions
with $J_1 = J_3 > 0$ (we will assume $J_2 > 0$
throughout this section) \cite{TKS}:
That of the $S=1/2$-$S=1$ ferrimagnetic chain for $J_2 < 0.90816 J_1$,
a spontaneously dimerized state for $0.90816 J_1 < J_2 < 2 J_1$
and, finally, singlets are formed on all bonds coupled by $J_2$ for
$J_2 > 2 J_1$ with effectively free spins in between. In the first
and the third case, one finds ferrimagnetic behavior with a spontaneous
magnetization $\Mag = 1/3$ and only the second region exhibits the
requested gap.

For $J_1 = J_3 < 0$, we found only two regions:
\begin{enumerate}
\item
For
\beq
J_2 > -J_1
\eeq{J2cF1}
the groundstate is formed by singlets on the $J_2$-bonds
and free $S=1/2$ spins in between. This again gives rise to ferrimagnetic
behavior with a spontaneous magnetization $\Mag = 1/3$.
\item
When
\beq
J_2 < -J_1
\eeq{J2cF2}
the entire system behaves like a ferromagnet. In this case
the system is spontaneously completely polarized ($\Mag=1$).
\end{enumerate}
We conclude that --unlike for $J_1 = J_3 > 0$--
the groundstate is always gapless for $J_1 = J_3 < 0$
which we have argued to be more
appropriate for Cu$_3$Cl$_6$(H$_2$O)$_2 \cdot$2H$_8$C$_4$SO$_2$.

\section{Effective Hamiltonians for the groundstate}

\label{secEffHam}

The low-energy behavior of the model Hamiltonian (\ref{Htrimer})
can be analyzed further using degenerate perturbation theory.
Truncation at a certain order of the coupling constants leads to effective
Hamiltonians which in some cases turn out to be well-known models.

We will use the abbreviations
\beq
\Jt{i} = {J_i \over J_1} \, , \qquad
\Jb{i} = {J_i \over J_2} \, .
\eeq{Jab}

\subsection{$J_1$ large and antiferromagnetic}

To test the method, we first consider the case of antiferromagnetic
$J_1$. For this purpose
we extend the first-order effective Hamiltonian of \cite{TOHTK}
for the case $J_1 \gg J_2$, $J_3 \ge 0$ to second order.
For $J_1$ large and antiferromagnetic, the groundstate-space of a
trimer is given by an $S=1/2$ representation. In this subspace of doublets,
the effective Hamiltonian has the form of a $J_1$-$J_2$ chain when
truncated after the second order:
\beq
H_{\rm eff.} =
{\cal J}_1 \sum_{i} \Sv_i \cdot \Sv_{i+1}
+ {\cal J}_2 \sum_{i} \Sv_i \cdot \Sv_{i+2} \, .
\eeq{Heff}
Here, the $\Sv_i$ are effective spin-1/2 operators.
The effective exchange constants are
\beq
{{\cal J}_1 \over J_1} = \frac{4}{9} \left( -\Jt{3} + \Jt{2}\right)
 -\frac{79}{405}\Jt{3}^2 + \frac{8}{135}\Jt{2}\Jt{3}
    + \frac{211}{1620}\Jt{2}^2
\eeq{J1eff}
and
\beq
{{\cal J}_2 \over J_1} = -\frac{91}{486}\Jt{3}^2 + \frac{22}{243}\Jt{2}\Jt{3}
            + \frac{10}{243}\Jt{2}^2 \, .
\eeq{J2eff}
In this approximation, the ferrimagnetic phase found in \cite{OTTK}
is given by an effective
ferromagnetic Hamiltonian (${\cal J}_1 < 0$) while the antiferromagnetic
phase corresponds to ${\cal J}_1 > 0$. The transition line can thus be
determined from ${\cal J}_1 = 0$ \cite{footnote4}. We find
\bea
\Jt{3} &=&
{\frac {12}{79}} \Jt{2}-{\frac {90}{79}}+{\frac {1}{158}}\sqrt{
17245 \Jt{2}^{2}+48240 \Jt{2}+32400} \nn \\
&=&  \Jt{2} - {1 \over 80} \Jt{2}^2 + {\cal O}\left(\Jt{2}^3\right) \, ,
\label{Jt3FD}
\eea
which improves the agreement of the approximation $\Jt{3} = \Jt{2}$
\cite{TOHTK} with the numerical results of \cite{OTTK}.

The dimer phase with a spin gap is characterized by
${\cal J}_2/{\cal J}_1 > 0.241167(5)$ (see \cite{Eggert} and
references therein). Using (\ref{J1eff}) and (\ref{J2eff}), it
is found to open at $\Jt{2} \approx 3.60$, $\Jt{3} \approx 1.361$ with
a square-root like behavior of $\Jt{3}$ as a function of $\Jt{2}$.
Since this is not in the weak-coupling region, it is not surprising
that the numbers differ substantially from those obtained numerically
in \cite{OTTK}. However, the topology of the groundstate phase
diagram comes out correctly from our effective Hamiltonian: In particular,
the dimerized spin gap phase is located inside the antiferromagnetic
phase and arises because of a sufficiently large effective second neighbor
frustration ${\cal J}_2$.

\subsection{$J_2$ large and antiferromagnetic}

The preceding argumentation is not applicable to the region $J_2 > 0$,
$J_1,J_3 < 0$. However, a similar case has been discussed earlier
\cite{Hida,poly} and $J_2 \gg \abs{J_1}, \abs{J_3}$
has been found to be a useful limiting case. We will now analyse
this region in the same manner as above.

For $J_2 \gg \abs{J_1}, \abs{J_3}$, the spins on all $J_2$-bonds couple to
singlets and only the intermediate spins contribute to the low-energy
excitations. In the space of these intermediate spins,
we can again map the Hamiltonian (\ref{Htrimer})
to the Hamiltonian (\ref{Heff}) to the lowest orders in $J_1$, $J_3$.
Up to fifth order, we find the effective coupling constants to be
given by \cite{footnote5}
\bea
{{\cal J}_1 \over J_2} &=& \left(\Jb{1}-\Jb{3}\right )^{2}
\left\{
{\frac {1}{2}}
+{\frac {3 \left (\Jb{1}+\Jb{3}\right )}{4}}
+3 \Jb{1} \Jb{3} \right. \nn \\
&& \left . -{\frac {
\left (\Jb{1}+\Jb{3}\right )\left (107 \left({\Jb{1}}^{2}+{\Jb{3}}^{2}\right )
-406 \Jb{1} \Jb{3} \right )}{64}}
\right\}
\label{J1effL2}
\eea
and
\beq
{{\cal J}_2 \over J_2} =
{\frac {\left (\Jb{1}+\Jb{3}\right )\left (\Jb{1}-\Jb{3}\right )^{4}}{4}} \, .
\eeq{J2effL2}
This mapping is now applicable regardless of the sign of $J_1$ and
$J_3$ as long as $J_2 > 0$. First we consider the case of
antiferromagnetic $J_1, J_3 > 0$. Then the effective coupling constants
are essentially always antiferromagnetic, {\it i.e.}\
${\cal J}_1, {\cal J}_2 > 0$
leading to a frustrated chain. If $J_1$ and $J_3$ are large enough,
${\cal J}_2 / {\cal J}_1$ can exceed the critical value of about
$0.241$ (see above) and a spin gap opens. These observations are again
in qualitative agreement with the phase diagram of \cite{OTTK}.
As for the preceding limit, one should not expect good quantitative
agreement since the required values of $J_1$ and $J_3$ are not small but
of the same order as $J_2$.

Now we turn to the more interesting case $J_1,J_3 < 0$.
Then the oupling constant (\ref{J2effL2}) is always ferromagnetic:
${\cal J}_2 < 0$. If $\abs{J_1}$ and $\abs{J_3}$ are large enough,
${\cal J}_1$ also becomes ferromagnetic. This is compatible
with the behavior found in section \ref{secTheLine}
on the line $J_1 = J_3 < 0$. If $\abs{J_1}$ and $\abs{J_3}$ are small,
${\cal J}_1$ remains antiferromagnetic. Since ${\cal J}_2$
is always ferromagnetic, no frustration arises in the
effective model and a spin gap is {\it not} expected
to open. This is true to the order which we have considered. Higher
orders might actually yield frustrating contributions. In any case,
frustration is substantially weaker for ferromagnetic $J_1,J_3 < 0$
than for antiferromagnetic $J_1,J_3 > 0$. It is therefore plausible
that a spin gap is absent in the ferromagnetic region (unless $\abs{J_1}$
and/or $\abs{J_3}$ are very large and the present argument is not
applicable).

It should be noted that (\ref{J1effL2}) and (\ref{J2effL2}) turn out
to be small if $\Jb1 - \Jb3$ is small. In fact, one can argue that
the results of this section remain qualitatively correct for
$\Jb1 - \Jb3$ small even if $\Jb1$ and $\Jb3$ are not separately small:
For $J_1 = J_3$, the intermediate spins are effectively decoupled due
to the presence of the singlets on the $J_2$-bonds (see section
\ref{secTheLine}).
A small detuning $J_1 \ne J_3$ generates an effective coupling of the
intermediate spins via higher-order processes.
However, the effective coupling will stay small as long as $J_1 - J_3$
is small. If one wants to model Cu$_3$Cl$_6$(H$_2$O)$_2\cdot$2H$_8$C$_4$SO$_2$,
$\abs{J_1 - J_3}$ must therefore at least be on the same scale
as e.g.\ the field $h \approx 80$K required to polarize the
intermediate spins leading to $\Mag = 1/3$ \cite{ITHUOTKN}. This
observation rules out a $J_1$ very close to $J_3$.

Finally, we also calculated the effective Hamiltonian for a strong ferromagnetic
intra-trimer interaction $J_1$. The problem then maps to a frustrated
$S=3/2$ chain with four-spin interactions. Even if this is not a well-known
Hamiltonian and the issue of a spin gap thus remains unclear in this case,
we present it in appendix \ref{appS3o2} in order to open the way for further
investigation of this limit.

\section{Magnetization plateaux}

\label{secMplateau}

We complete our theoretical analysis with a discussion of
plateaux in the magnetization curves of the frustrated trimer
chain model.

A plateau with $\langle M \rangle = 1/3$ is abundant in the
magnetization curve (compare Fig.\ \ref{mCurve}) and can be easily
understood in the limits $\abs{J_2},\abs{J_3} \ll J_1$ or
$\abs{J_1},\abs{J_3} \ll J_2$. This is readily done by adding the coupling
$J_3$ to the series of \cite{poly}. More details as well as the
explicit series for the boundaries of the $\langle M \rangle = 1/3$ plateau
are available under \cite{WWW}. Here we just mention that the main
conclusions of \cite{poly} regarding this plateau remain qualitatively
unchanged in the presence of the additional coupling $J_3$. 

Regarding plateaux with $\langle M \rangle \ne 1/3$, observe first that,
when a spin gap opens in the frustrated trimer model, the groundstate
is dimerized, {\it i.e.}\ translational invariance is spontaneously broken
by a period two. Spontaneous breaking of translational invariance by a period
two also permits the appearance of a plateau with $\langle M \rangle=2/3$
(see \cite{zigzag} and references therein). We will now investigate
this possibility further.

First we consider the case $J_1 > 0$ and
start in the limit of strong trimerization ($J_2=0,J_3=0$).
When one applies a magnetic field $h_c=\frac{3}{2}J_1$, the two
states $\left|\uparrow\uparrow\uparrow\rangle\right.$ and
$\frac{1}{\sqrt6}(\left|\downarrow\uparrow\uparrow\rangle\right.
-2\left|\uparrow\downarrow\uparrow\rangle\right.
+\left|\uparrow\uparrow\downarrow\rangle\right.)$ are
degenerate in energy. This degeneracy is then lifted by the couplings
$J_2,J_3$. The effective Hamiltonian to first order is an $XXZ$ chain in a
magnetic field \cite{Totsuka,Mila,CJYFHBLHP,TLPRS,FuZh,WeHa,zigzag}.
We obtain the following effective couplings for the $XXZ$ chain:
\bea
J_{xy}&=&\frac{1}{6}J_2-\frac{2}{3}J_3 \nonumber \\
J_{z}&=&\frac{1}{36}(J_2+8 J_3) \nonumber \\
h_{{\rm eff}}&=&h-h_c-\frac{1}{36}(5 J_2 +22 J_3)
\label{effXXZ}
\eea
and therefore the effective anisotropy $\Delta_{{\rm eff}} = J_{z}
/\abs{J_{xy}}$ is
\beq
\Delta_{{\rm eff}}=\frac{J_2+8 J_3}{\abs{6 J_2-24 J_3}} \, .
\eeq{aniso}
For $5/32 < J_3/J_2 < 7/16$, we have $\Delta_{{\rm eff}} > 1$
and thus a gap, {\it i.e.}\ an $\langle M \rangle = 2/3$ plateau
in the original model. A plateau with $\langle M \rangle = 2/3$
can be indeed observed numerically somewhere in this region
(see e.g.\ \cite{TOHTK}).
The line $J_3/J_2=1/4$ describes the Ising limit $\Delta_{{\rm eff}} = \infty$.

In order to address the region of ferromagnetic $J_1$, we now
start from the limit $J_1 = J_3 = 0$ and apply a magnetic
field $h_c = J_2$. Then the two states $\left|\uparrow\uparrow\rangle\right.$
and ${1 \over \sqrt{2}}\left(\left|\downarrow\uparrow\rangle\right. -
\left|\uparrow\downarrow\rangle\right.\right)$ on the $J_2$-dimer become
degenerate in energy while the intermediate spins are already polarized.
This can be again treated by degenerate perturbation theory in $1/J_2$.
Up to third order we find an $XXZ$ chain with
\bea
{J_{xy} \over J_2} &=& {1 \over 8}  \left(2+\Jb1+\Jb3\right)
            \left(\Jb1-\Jb3\right)^2 \nonumber \\
{J_{z} \over J_2} &=& {1 \over 8}  \left(\Jb1+\Jb3\right)
            \left(\Jb1-\Jb3\right)^2 \nonumber \\
{h_{\rm eff} \over J_2} &=& {h \over J_2}
  - 1 - {1 \over 2} \left(\Jb1+\Jb3\right)
      - {1 \over 4} \left(\Jb1-\Jb3\right)^2
\label{effXXZ2}
\eea
that is
\beq
\Delta_{\rm eff} = {\Jb1+\Jb3 \over \abs{2+\Jb1+\Jb3}} \, .
\eeq{aniso2}
In the region where this treatment is valid, we always have
a small $\Delta_{\rm eff}$, {\it i.e.}\ no plateau at $\langle M \rangle=2/3$.
Indeed, one can see that the dimer excitations can hop at second order
in $1/J_2$ while up to this order all diagonal terms involve only
a single dimer site. Thus, up to second order the diagonal terms
contribute only to $h_{\rm eff}$ and to this order one obtains an
$XY$ chain in a magnetic field. A small anisotropy is restored at
third order before terms that are not described by a simple $XXZ$ chain
arise at fourth order.

\section{Conclusions}

We have studied the frustrated trimer chain (\ref{Htrimer}) (Fig.~\ref{Model})
using a variety of methods. First, we have computed 12th-order
high-temperature series for the susceptibility $\chi$ and specific heat.
Fits of the high-temperate tail of the susceptibility computed from the
model to the one measured on Cu$_3$Cl$_6$(H$_2$O)$_2\cdot$2H$_8$C$_4$SO$_2$
\cite{ITHUOTKN} lead to $J_2 = 250{\rm K} \pm 40{\rm K}$ and
{\it ferromagnetic} $J_1 = -260{\rm K} \pm 50{\rm K}$,
$J_3 = -40{\rm K} \pm 30{\rm K}$ (we showed in appendix \ref{appAF} that
$\chi(T)$ cannot be fitted with the antiferromagnetic parameters proposed in
\cite{OTTK,SaTa,TOHTK}).
We assumed that these parameters remain valid down to low temperatures
since we are not aware of any indication of a drastic change in
the magnetic behavior of Cu$_3$Cl$_6$(H$_2$O)$_2\cdot$2H$_8$C$_4$SO$_2$
as temperature is lowered. In fact, features of other experimental
observations at intermediate and low temperatures are roughly reproduced
with the aforementioned parameters: We find a maximum in $\chi(T)$ in the
region $50{\rm K} \le T \le 100{\rm K}$ and a smooth increase of
the low-temperature magnetization $\langle M \rangle$ from $0$
to $1/3$ as the external magnetic field is increased from zero
to several ten Tesla. From a quantitative point of view, the agreement
may however not yet be entirely satisfactory: Deviations between the measured
susceptibility from the one obtained within the model can be seen
in the interval $80{\rm K} \le T \le 200{\rm K}$ and the model
predicts an $\Mag = 1/3$ magnetization for a magnetic field that
is a factor two to three below the one actually required in the
experiment.

Probably
the most exciting experimental observation \cite{ITHUOTKN} for
Cu$_3$Cl$_6$(H$_2$O)$_2\cdot$2H$_8$C$_4$SO$_2$ is the existence of
a spin gap of about 5.5K. We have therefore searched for a spin gap
in the region of ferromagnetic $J_1$ and $J_3$ using several methods.
Neither Lanczos diagonalization, discussion of the line $J_1 = J_3$
nor an effective Hamiltonian for large $J_2$ provide any evidence
in favor of a spin gap in this parameter region. A further careful
analysis of this issue would certainly be desireable in particular
in view of the small size of the actually observed gap. At present,
however, it seems likely that the model does not reproduce a spin
gap in the relevant parameter region.

It should be noted that the coupling constants which we have determined
are about two orders of magnitude larger than the experimentally
observed gap. Therefore, a small modification of the model is
sufficient to produce a gap of this magnitude.
The possibilities include dimerization of the coupling constants,
exchange anisotropy as well as additional couplings. A modification
of the model along these lines may also help to improve the quantitative
agreement with the features observed in
Cu$_3$Cl$_6$(H$_2$O)$_2\cdot$2H$_8$C$_4$SO$_2$
at energy scales of about 100K. Further measurements are however needed
to discriminate between these possibilities. For example, it would
be interesting to measure the specific heat and compare it with
our series (\ref{cvHT}). It emerges also from our analysis that a
temperature of 300K is still too small to allow for application of
a simple Curie-Weiss law to the magnetic susceptibility $\chi$. It would
therefore be useful to measure $\chi$ to higher temperatures in order to permit
analysis via truncation of (\ref{chiHT}) after the order $T^{-2}$
which would provide a more direct check that $2(J_1+J_3)+J_2$ is
negative. 

However, inelastic neutron scattering would presumably
be most helpful: First, this should clearly decide if
Cu$_3$Cl$_6$(H$_2$O)$_2\cdot$2H$_8$C$_4$SO$_2$ is really quasi-one-dimensional
and secondly it would yield direct information on the excitation spectrum
which could hopefully be interpreted in terms of coupling constants.
Such a determination of the coupling constants would also circumvent the
question whether model parameters change as a function of temperature since
neutron scattering is carried out at low temperatures, {\it i.e.}\ the
temperature scale of interest. We therefore hope that neutron scattering can
indeed be performed and are curious if excitations will be observed that are
similar to those computed in the trimer chain model (Fig.\ \ref{dispFig}).

The frustrated trimer chain model is also interesting in its own right:
It has a rich phase diagram which among others includes many aspects of the
$J_1$-$J_2$ chain such as a frustration-induced spin gap in some parameter
region \cite{OTTK,SaTa,TOHTK}. Also plateaux in the magnetization
curve exist in this model: A plateau with $\Mag = 1/3$ is abundant both
in the regions with antiferromagnetic and ferrimagnetic $h=0$ groundstates.
Also a plateau with $\Mag = 2/3$ can be shown to exist in
the region with $J_1, J_3 > 0$ (see \cite{TOHTK} and section \ref{secMplateau}).
Like in the case of the spin gap, the opening of the latter plateau is
accompanied by spontaneous breaking of translational invariance in the
groundstate. Amusingly, however, the $\Mag = 2/3$ plateau opens already for
$J_2, J_3 \ll J_1$ -- a region where the spin gap is absent.
In this context of magnetization plateaux, we hope that the magnetization
measurements \cite{ITHUOTKN} can be extended to slightly higher fields
which should unveil the lower edge of the $\Mag = 1/3$ plateau.

\section*{Acknowledgments}

We are very grateful to M.\ Ishii and H.\ Tanaka for providing us with their
partially unpublished data for the susceptibility and for discussions. In
addition, useful discussions with D.C.\ Cabra, F.\ Mila, M.\ Troyer and
T.M.\ Rice are gratefully acknowledged. A.H.\ is indebted to the Alexander
von Humboldt-foundation for financial support during the initial stages
of this work as well as to the ITP-ETHZ for hospitality.

\appendix
\section{Antiferromagnetic coupling constants}

\label{appAF}

In this appendix we discuss a fit of the magnetic susceptibility $\chi$
with antiferromagnetic coupling constants $J_i > 0$. A number of
assumptions are necessary in order
to obtain at all a convergent fit with parameters in the
antiferromagentic spin gap region \cite{OTTK,SaTa,TOHTK}.

First we fix the ratio of the magnetic field $h(\langle M \rangle = 1/3)$
to the spin gap $h_c(\langle M \rangle = 0)$ approximately to the
experimental value \cite{ITHUOTKN}
\beq
{h(\langle M \rangle = 1/3) \over h_c(\langle M \rangle = 0)}
= 14.1 \, .
\eeq{chooseRat}
To this end, we used numerical data for $h(\langle M \rangle = 1/3)$
and $h_c(\langle M \rangle = 0)$ on systems of size $L=12$, $18$ and $24$.
This data was extrapolated to $L=\infty$ in the same manner as in \cite{SaTa},
{\it i.e.}\ with a polynomial fit $h_L(\langle M \rangle) =
h_\infty(\langle M \rangle) + a/L + b/L^2$. For the spin gap, this
amounts to reproducing the computation of \cite{SaTa}.
The numerical solutions to eq.\ (\ref{chooseRat}) were then approximated
by
\beq
\Jt{3} = 0.3 \left(\Jt{2}-0.85\right)^2 + 0.63
\eeq{approxAF}
where we used the notation (\ref{Jab}). An analytic formula was needed
in order to implement the constraint (\ref{chooseRat}) by inserting
(\ref{approxAF}) into (\ref{chiHT}) before performing a fit.
Eq.\ (\ref{approxAF}) is valid for $0.6 \lesssim \Jt{2} \lesssim 2$.

\begin{figure}
\centerline{\psfig{figure=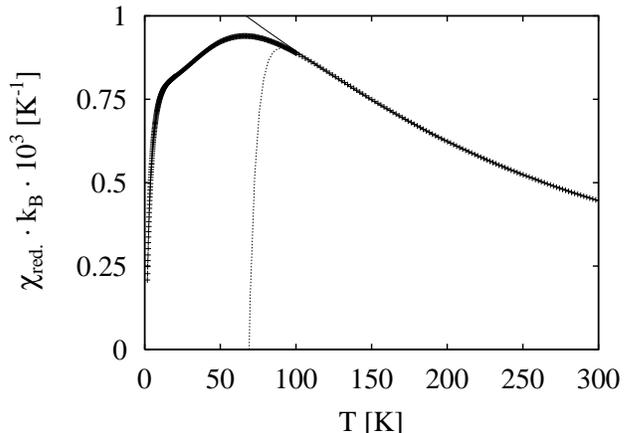,width=\linewidth,angle=270}}
\caption{
Experimental results for the susceptibility (`$+$') in comparison
with the fit $J_1 = 120$K, $J_2 = 141$K and $J_3 = 79$K.
We show the raw 12th order series (dotted line) as well as the
[7,6] Pad\'e approximant (full line).
\label{chiFigAF}
}
\end{figure}

The constraint (\ref{approxAF}) is still not sufficient to ensure
antiferromagnetic $J_i > 0$ with $0.6 \le \Jt{2} \le 2$. To achieve
this goal, we had to make the following further adjustments when fitting
our series (\ref{chiHT}) to the experimental data \cite{ITHUOTKN}:
\begin{enumerate}
\item Keep $g$ as a fitting parameter,
\item add a constant to (\ref{chiHT}) and use this as another parameter
      in the fit,
\item start fitting at low temperatures $T_l \approx 100$K.
\end{enumerate}
Note that both $g$ and
the additive constant turn out to be quite large. For example, for the
parameters used in Fig.\ \ref{chiFigAF}, we found $g \approx 2.9$ 
and an additive constant of about $-0.17 \cdot 10^{-3} {\rm K}^{-1} /k_B$.
This means that the prefactor in (\ref{chiExp}) is off by a factor of
about 2 from the value determined by ESR and that the absolute value of
the additive constant is almost 40\% of the susceptibility observed
at $T= 300$K !

On the basis of these unrealistic parameters, one could
already discard this fit to $\chi(T)$. Nevertheless, we compare it
to the one shown in Fig.\ \ref{chiFig}: Fig.\ \ref{chiFigAF} shows the
measured susceptibility for the polycrystalline sample \cite{ITHUOTKN}
together with the series evaluated at $J_1 = 120$K, $J_2 = 141$K and
$J_3 = 79$K. This parameter set
is close to parameters proposed in \cite{SaTa}. This proposal was based
on two assumptions: 1) The model should give rise to the experimentally
observed spin gap of around 5K \cite{ITHUOTKN}. 2) The maximum of $\chi$
is located at $T \approx 0.7 J_1$. While we do indeed reproduce the
spin gap rather accurately, the second assumption is
falsified by our computation: The frustration pushes the maximum
of $\chi(T)$ again to lower temperatures as compared to a non-frustrated
system. 

Fig.\ \ref{chiFigAF} should be compared to Fig.\ \ref{chiFig}. The seemingly
better agreement in the region $100{\rm K} \le T \le 200$K is due to including
this temperature interval in the fit for Fig.\ \ref{chiFigAF}, but not
in Fig.\ \ref{chiFig}. Note that the $\abs{J_i}$ are now
smaller by a factor of about two than those used in Fig.\ \ref{chiFig}.
One would therefore expect better convergence in the vicinity of the
maximum of $\chi(T)$, {\it i.e.}\ for $50 {\rm K} \le T \le 100$K. This
expectation is confirmed by the fact that in Fig.\ \ref{chiFigAF},
the [7,6], [6,6] and [6,5] Pad\'e approximants are indistinguishable.
However, while the series reproduces the maximum roughly in
Fig.\ \ref{chiFig}, this is definitely not the case in Fig.\ \ref{chiFigAF}.
The better agreement of the fit in Fig.\ \ref{chiFig} with the experimental
data at $T\approx 70$K is particularly remarkable since this temperature range
is far from the fitting region in this case, while closeby temperatures were
used in Fig.\ \ref{chiFigAF}.
In combination with the unrealistic assumptions needed to obtain a convergent
fit with all $J_i > 0$ one can therefore conclude safely that only
antiferromagnet coupling constants are not suitable for describing
the experimental data \cite{ITHUOTKN} for the susceptibility $\chi(T)$.

\section{Effective Hamiltonian for $J_1$ large and ferromagnetic}

\label{appS3o2}

For a strong ferromagnetic intra-trimer
interaction $J_1$, the noninteracting groundstates are built from
products of trimer $S=3/2$ states. Up to second order we find the
following effective Hamiltonian in this subspace of low-lying trimer quartets:
\bea
H_{\rm eff.} &=& \Jc{a} \sum_{i} \Sv_{i} \cdot \Sv_{i + 1}
 +\Jc{b} \sum_{i}  \Sv_{i} \cdot \Sv_{i + 2} \nn \\
&& +\Jc{c} \sum_{i}  \left(\Sv_{i} \cdot \Sv_{i + 1}\right)^2 \nn \\
&& +\frac{\Jc{d}}{2} \sum_{i} \bigl\{ (\Sv_{i} \cdot \Sv_{i + 1})
    (\Sv_{i+1} \cdot \Sv_{i + 2}) \nn \\
&& \qquad + (\Sv_{i+2} \cdot \Sv_{i + 1})(\Sv_{i+1} \cdot \Sv_{i}) \bigr\} \, ,
\label{Heff32}
\eea
where the $\Sv_{i}$ are now effective spin-3/2 operators.

The coupling constants are found to be:
\bea
\Jc{a}&=&\frac{1}{9}(J_2+2J_3)
+ \frac{197 J_2^2 + 212 J_2 J_3 + 212 J_3^2}{2592 \abs{J_1}} \, , \nn\\
\Jc{b}&=&\frac{2J_2^2 + 5 J_2 J_3 + 2 J_3^2}{27 \abs{J_1}} \, ,\nn\\
\Jc{c}&=&\frac{41J_2^2 + 100 J_2 J_3 + 36 J_3^2}{1296 \abs{J_1}} \, , \nn\\
\Jc{d}&=&-\frac{4(2 J_2^2 + 5 J_2 J_3 + 2 J_3^2)}{243 \abs{J_1}} \, .
\label{J3o2eff}
\eea
Even if this effective Hamiltonian is not a well-known one,
it is clear that there is no spin gap in first order, since
then the system is effectively a nearest neighbor $S=3/2$ Heisenberg chain
which is either gapless ($J_2+2J_3>0$) or ferromagnetic ($J_2+2J_3<0$).

If one neglects the $\Jc{c}$ and $\Jc{d}$ terms, one obtains
a frustrated $S=3/2$ chain which has been investigated with DMRG
and leads to a gap for $\Jc{b}/\Jc{a} \gtrsim 0.3$ \cite{RSC}.
It seems to be possible to obtain antiferromagnetic $\Jc{a}$ and
$\Jc{b}$ in this region if $J_2$ and $J_3$ are chosen suitably
and large (a region including the coupling constants determined in
section \ref{secChiFit}). However, then one is not in the perturbative region
anymore and the $\Jc{c}$ and $\Jc{d}$ terms may also become important.
Further discussion is therefore needed for reliable conclusions
about a gap on the basis of the Hamiltonian (\ref{Heff32}) with
coupling constants (\ref{J3o2eff}).

\end{multicols}


\begin{thebibliography}{99}
\bibitem{footnote1} The low-temperature susceptibility of \cite{ITHUOTKN} is
  substantially lower than that of \cite{SLW}. The authors of
  \cite{ITHUOTKN} suggested that this discrepancy may be due to a large
  impurity contribution in \cite{SLW}.
\bibitem{footnote2}
  We use the notations of \cite{OTTK} which differ from \cite{ITHUOTKN}
  by exchanging $J_2$ and $J_3$.
\bibitem{footnote3} The price to pay for using an elementary approach is
   a substantial amount of CPU time even on modern workstations.
\bibitem{footnote4} ${\cal J}_2$ changes sign inside the region with
       ${\cal J}_1 > 0$ such that ${\cal J}_2 < 0$ throughout the region
       with ${\cal J}_1 < 0$. ${\cal J}_2$ is therefore not relevant to the
       transition.
\end{thebibliography}
\end{document}